# *The solid-liquid transition in complex fluids*

Reinhard Höhler, Sylvie Cohen-Addad, Vincent Labiausse

Université de Marne-la-Vallée, Laboratoire de Physique des Matériaux Divisés et des Interfaces, UMR CNRS 8108, 5 Bd Descartes, 77454 Marne-la-Vallée cedex 2, France

*The yielding of foams, concentrated emulsions, pastes and other soft materials under applied strain is often characterized by measuring the complex shear modulus as a function of strain amplitude at low frequency. Results obtained for materials of different physico-chemical constitution are strikingly similar, suggesting the presence of a generic mechanism as pointed out in several previous theoretical studies. We show that the observed strain dependence of the complex shear modulus can be interpreted as the consequence of elasto-plastic response.*

## I. Introduction

Many forms of disordered soft condensed matter such as foams, emulsions and soft pastes show strikingly similar rheological behaviour. Among its principal features are slow relaxations, and a transition between viscoelastic solid-like and Non-Newtonian liquid-like response as a function of applied stress or strain (Barrat et al 2003). The liquid-solid transition is often probed by measuring the stress response to an applied oscillatory strain at low frequency and analysed in terms of a strain amplitude dependent complex shear modulus $G^*(\gamma_0) = G'(\gamma_0)+iG''(\gamma_0)$ (Derec et al 2003; Mason et al 1995; Saint-Jalmes & Durian 1999). At large strain amplitudes $\gamma_0$, the viscoelastic response is non-linear, giving rise to a stress response that contains a spectrum of harmonics. In this case $G^*(\gamma_0)$ is usually calculated on the basis of the fundamental component of this spectrum and therefore provides a characterization of the behaviour which is still well defined but not complete. Figure 1 shows the complex shear modulus of a foam, an emulsion and a paste whose linear viscoelastic loss modulus G'' is much smaller than the storage modulus G'. Detailed sample characteristics are specified in the appendix. The data are indeed very similar in the chosen representation where $G^*(\gamma_0)$ is normalized by the value of G' in the linear regime and where the strain amplitude is normalized by the yield strain $\gamma_y$: G' is constant at small strain amplitudes $\gamma_0$, starts decreasing at the yield strain and finally drops as $\gamma_0^{-1.5}$ at large strain amplitudes. G'' presents a plateau at low strain amplitudes. With increasing amplitude, it goes through a maximum close to the yield strain and then drops asymptotically as $\gamma_0^{-1}$. The similar response observed for materials of different physico-chemical constitutions has stimulated the development of various minimal generic models (Derec et al 2003; Hébraud & Lequeux 1998; Miyazaki et al 2006; Sollich et al 1997). In the Soft Glassy Rheology (SGR) model, a complex fluid is described as an ensemble of mesoscopic regions. Their response is supposed to be linearly elastic up to a maximum local elastic energy E, randomly chosen from a distribution ρ(E) (Sollich 1998; Sollich et al 1997). If this threshold is exceeded, a structural rearrangement occurs, all the elastic energy is dissipated and a new value of E is chosen. The distribution of the energies E can also be interpreted in terms of a local yield stress. Moreover, according to this model rearrangements can also be activated by strain fluctuations, represented in a mean field approach by an effective noise temperature ζ. It has been shown that the choice ρ(E) ∝ exp(-E/ ζ) with a specific value for ζ leads to a rheological constitutive equation that qualitatively describes many features of the behaviour observed for soft yield stress materials, and in particular the behaviour shown on figure 1 is predicted at least approximately. Another group of models are based on a scalar measure of the "degree of jamming" or structural relaxation time, which is coupled to the flow history of the material: flow breaks up the jammed structure and reduces the viscosity. In contrast, aging re-establishes the jammed structure and enhances the viscosity. This is expressed in terms of non-linear differential equations containing free parameters that reproduce at least qualitatively much of the experimentally observed phenomenology (Coussot et al 2002; Derec et al 2003; Miyazaki et al 2006; Picard et al 2002). Several mode coupling theories have also been proposed (Hébraud & Lequeux 1998; Miyazaki et al 2006).

In this paper, we point out that the seemingly complex phenomenology shown on Figure 1 can be explained on a very simple basis: The materials that we consider all have in common that they behave mainly elastically at small strain and that they present plastic flow beyond a yield strain. It is therefore natural to consider elementary elasto-plastic behaviour, mimicked by a spring connected in series with a slider, as a simple starting point for a constitutive model. We will show that such an elementary elasto-plastic description is sufficient to describe accurately the normalized data shown in figure 1, without any free fit parameters for complex yield stress fluids that are weakly dissipative in the linear viscoelastic regime.

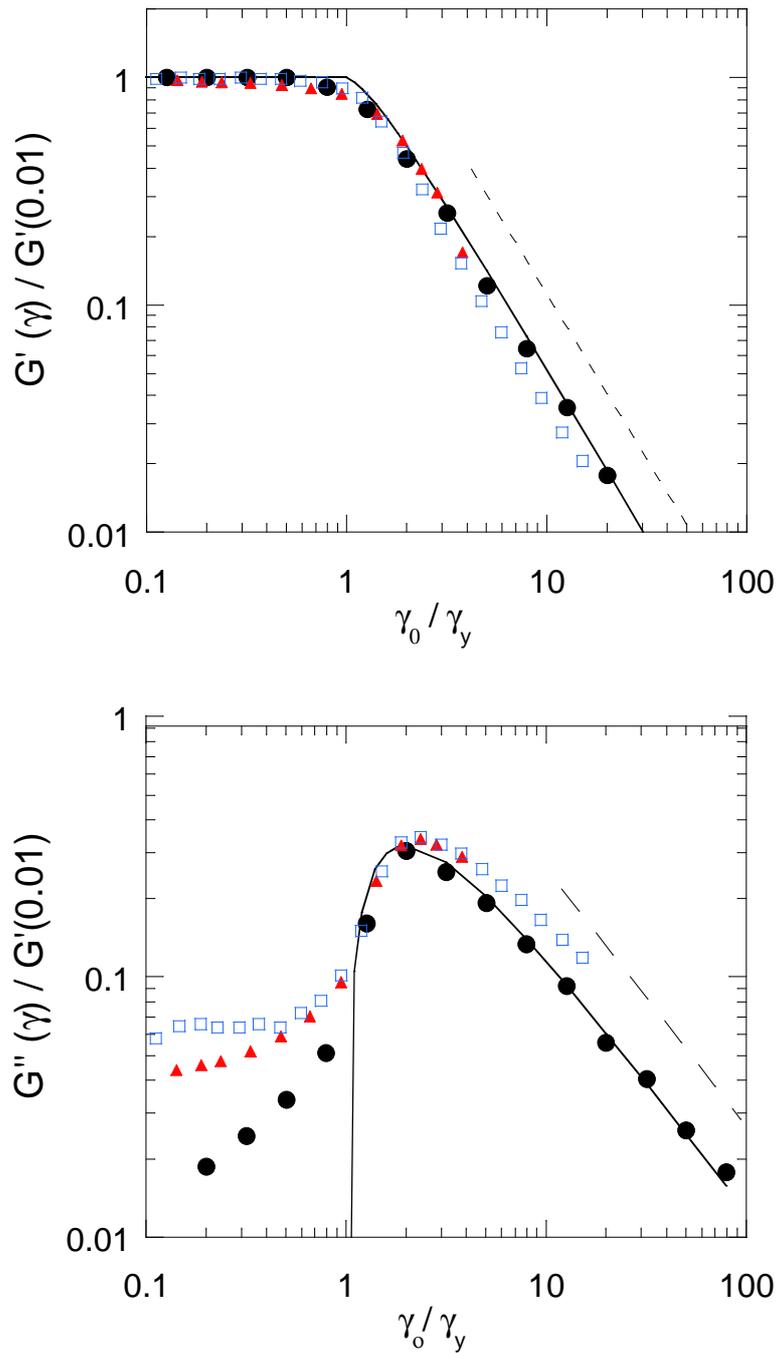

Figure 1 Real and imaginary parts of the complex shear modulus as a function of strain amplitude $\gamma_0$ at low frequency. The moduli are normalised by the value of G' in the linear viscoelastic regime, $\gamma_0$ is normalized by the amplitude where yielding sets in. The full line is the prediction of Eq 2, the symbols correspond to experimental data: □: a soft paste (Derec et al 2003), ▲ a dry aqueous foam, ● a concentrated emulsion, (Mason et al 1995). The detailed sample characteristics are presented in the appendix. The dashed lines (slopes -1.5 in

*the G' plot and -1 in G" plot) help to appreciate the asymptotic power law behaviour of the data.*

## II. Oscillatory response of an ideal elasto-plastic material

The spring in the elasto-plastic model schematically shown in the figure 2a represents the elastic response observed at small stress. It is characterized by a shear modulus denoted as G. The slider behaves as rigid link when subjected to a force whose modulus is below a critical value and it moves freely for larger positive or negative forces. The force corresponds to the shear stress $\tau$ in the material and its critical value to the yield stress $\tau_y$. Plastic strain and elastic strain respectively correspond to elongations of the slider and of the spring. The total strain of the material that can be imposed experimentally is denoted $\gamma$ and given by the sum of its elastic and plastic components. The yield strain in the absence of plastic strain, denoted $\gamma_y$, is obtained by dividing the yield stress by G. We now calculate the response of such a model to an imposed strain $\gamma(t) = \gamma_0 \cos(\omega t)$. If $\gamma_0$ is smaller or equal to $\gamma_y$, the stationary response is obviously purely elastic and $G^* = G$. In this case, the stress oscillates sinusoidally with the same phase as the strain.

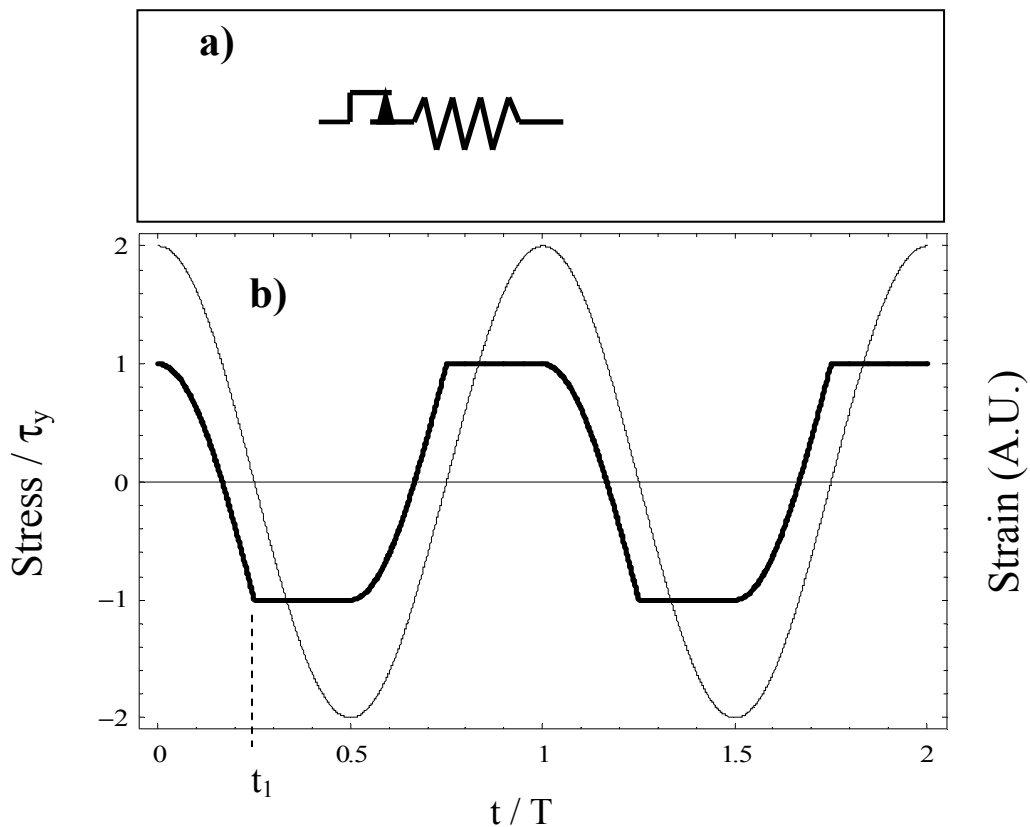

*Figure 2 a) Schematic representation of the elementary elasto-plastic model discussed in the text b) Time evolution of the stress (thick line) if an oscillatory strain (thin line) is imposed*

*whose amplitude exceeds the yield strain. The time is normalized by the period of the oscillation, denoted as T. The significance of the time denoted $t_1$ is explained in the text.*

We now discuss the regime $\gamma_0 > \gamma_y$ where the stress in the sample periodically reaches the yield stress so that the slider is triggered as illustrated by figure 2. Let us consider an instant where this happens at some point during the increase of the applied strain. From then on and until the next maximum of applied strain is reached, the slider remains mobile and the stress saturates at the value $\tau_y$. When the applied strain goes beyond the maximum, for instance at the time t = 0, the stress decreases and the slider becomes rigid again. From here on, the evolution of the applied strain induces a stress variation $\tau(t) = G \gamma_y + G \gamma_0 (\cos(\omega t) - 1)$ until the stress decreases to $-\tau_y$ at a time denoted as $t_1$. An elementary calculation yields

$$t_1 = arccos(-2\gamma_y / \gamma_0 + 1)/\omega \tag{1}$$

The stress remains saturated until the minimum of applied strain is reached at a time which is half the period of the oscillation $T = 2\pi/\omega$. As the strain starts increasing again, the stress increases as $\tau(t) = -G \gamma_y + G \gamma_0 (\cos(\omega t) + 1)$ until it reaches the value $\tau_y$, at the time $T/2 + t_1$. The evolution then becomes saturated until the time t = T, and from then on the whole process continues periodically. To determine $G^*(\gamma_0)$, we calculate the Fourier component of the stress oscillation at the frequency $\omega$ and divide it by the strain amplitude. These expressions are simplified without any loss of generality by scaling stress and strain such that G = 1 and $\gamma_y = 1$.

$$G^* = \frac{2}{\gamma_0 T}\left\{\int_0^{t_1}(1+\gamma_0(\cos(\omega t)-1))e^{-i\omega t}dt - \int_{t_1}^{T/2} e^{-i\omega t}dt + \int_{T/2}^{T/2+t_1}(-1+\gamma_0(\cos(\omega t)+1))e^{-i\omega t}dt + \int_{T/2+t_1}^{T} e^{-i\omega t}dt\right\}$$

(2)

$$= \frac{1}{\pi}\left(arccos\left(\frac{-2+\gamma_o}{\gamma_o}\right) - \frac{2(-2+\gamma_o)}{\gamma_o}\sqrt{\frac{-1+\gamma_o}{\gamma_o^2}} + i\frac{4(\gamma_o-1)}{\gamma_o^2}\right)$$

Experimental data can be compared to this expression by the normalization described in the introduction. An asymptotic development of Eq.(2) for large strain amplitudes $\gamma_0$ shows that G' and G" respectively scale as $\gamma_0^{-3/2}$ and $\gamma_0^{-1}$.

## III. Discussion

Figure 1 shows excellent agreement between the experimentally measured real part of the complex shear modulus of all three soft materials considered here and the elastoplastic model. In particular the observed asymptotic power law scaling with strain amplitude is correctly predicted by the model, without any adjustable parameters. Concerning the imaginary part of the shear modulus, the model correctly predicts the experimentally observed power law behaviour at large strain amplitudes. The agreement is excellent down to the yield strain for the emulsion whereas for the paste, the dissipation is somewhat larger than predicted. This is consistent with the behaviour at strain amplitudes smaller than the yield strain where the emulsion is closest to the model prediction G" = 0 whereas the paste and the foam show a larger value of G". This suggests that additional dissipation mechanisms present in the linear regime are also present when the material yields. Indeed, the elasto-plastic description is best adapted for materials which are weakly dissipative in the linear regime. For foams and emulsions, this is true for high dispersed volume fractions $\phi$ in particular if the rate of droplet or bubble rearrangements due to coarsening is small (Cohen-Addad et al 2004; Vincent-Bonnieu et al 2006). For $\phi$ approaching 0.64 these systems loose their rigidity and in the vicinity of this transition, the data published in the literature (Mason et al 1995; Saint-Jalmes & Durian 1999) show significant deviations from elastoplastic behaviour.

A useful extension of the elasto-plastic model would be to model dissipation in the linear viscoelastic regime as well as viscous effects that may come into play at the yield stress. For instance, foams have been shown to behave as Maxwell liquids (Cohen-Addad et al 2004) and replacing the elastic by such a viscoelastic response would help to make the model more general but this is beyond the scope of the present paper. Let us finally clarify a physical origin of the constituent elements of the elasto-plastic model. For foams and emulsions, Princen has shown that the elasticity (represented here by the spring) is due to the change of interfacial energy induced when the foam is sheared. A detailed analysis shows that G scales as the interfacial tension divided by an average bubble radius (Reviews are given in (Höhler & Cohen-Addad 2005; Weaire & Hutzler 1999)). The slider represents bubble rearrangements that arise when under the effect of an applied strain, bubble edges shrink to zero length, creating unstable configurations. This process has also been analyzed by Princen who showed that the yield stress scales as the yield stress divided by an average bubble radius (Höhler & Cohen-Addad 2005; Weaire & Hutzler 1999).

The analysis of the complex shear modulus as a function of strain amplitude in terms of an elasto-plastic response presented in this paper provides a surprisingly good description of foam, emulsions and pastes, without invoking the complex features and parameters of the models mentioned in the introduction. However, probing the fundamental Fourier component of the stress response may not be sufficient to distinguish the presence of a single yield stress from the distribution of yield stresses postulated in the SGR model or to detect the dynamics of fluidization and jamming modelled in the models cited in the introduction. Applying Fourier transform rheology experiments to soft yield stress materials could be a good approach for gaining deeper insight.

## IV. Conclusion

There is a striking similarity between the yielding behaviours of various soft solids that are weakly dissipative in the linear regime. Their response probed in oscillatory experiments can be explained accurately as a consequence of two simple features: elastic response at small strains and yielding beyond a threshold $\tau_y$. This suggests that elementary elastoplastic response may be a generic feature of weakly dissipative jammed soft materials.

## Acknowledgements

We would like to thank Florence Rouyer for stimulating discussions.

## Appendix

In this appendix, we present the characteristics of the samples represented in figure 1. The emulsion studied in the experiments by Mason et al. (Mason et al 1995) were comprised of stabilized silicon oil droplets in water whose radius was 0.53μm ± 10%. The data shown on figure 1 were obtained for an effective droplet volume fraction of 0.8 and at a frequency of 1 rad/s. The paste studied by Derec et al. (Derec et al 2003) was made of silica particles (diameter: 0.1μm) dispersed in water, stabilized to behave as soft repulsive colloidal spheres. These experiments were carried out at a frequency of 1 Hz. The foam data were measured by us at a frequency of 1 Hz using a Rheometer (Bohlin CVOR-150) with a cylindrical Couette cell. Its surfaces in contact with the sample were roughened to prevent wall slip. The foaming liquid was an aqueous solution containing an anionic surfactant (1.5% g/g AOK, Witco chemicals), dodecanol (0.2 g/g %) as well as polyethylene oxide (0.4 g/g). The sample gas volume fraction was 0.97, the average bubble size was 42 μm. The gas was nitrogen containing perfluorohexane vapour which is insoluble in water. This vapour slows down coarsening of the foam and therefore makes it more stable (Safouane et al 2001).

**References**

Barrat J-L, Feigelman M, Kurchan J, Dalibard J, eds. 2003. *Slow relaxations and nonequilibrium dynamics in condensed matter*, Vols. LXXVII. Berlin: Springer

Cohen-Addad S, Höhler R, Khidas Y. 2004. Origin of the Slow Linear Viscoelastic Response of Aqueous Foams. *Physical Review Letters* 93:028302-4

Coussot P, Nguyen QD, Huynh HT, Bonn D. 2002. Avalanche Behavior in Yield Stress Fluids. *Physical Review Letters* 88:175501-4

Derec C, Ducouret G, Ajdari A, Lequeux F. 2003. Aging and nonlinear rheology in suspensions of polyethylene oxide--protected silica particles. *Physical Review E* 67:061403-9

Hébraud P, Lequeux F. 1998. Mode-Coupling Theory for the Pasty Rheology of Soft Glassy Materials. *Physical Review Letters* 81:2934-7

Höhler R, Cohen-Addad S. 2005. Rheology of liquid foams. *Journal of Physics: Condensed Matter* 17:R1041-R69

Mason TG, Bibette J, Weitz DA. 1995. Elasticity of Compressed Emulsions. *Physical Review Letters* 75:2051-4

Miyazaki K, Wyss HM, Weitz DA, Reichman DR. 2006. Nonlinear viscoelasticity of metastable complex fluids. *Europhys. Lett.* 75:915-21

Picard G, Ajdari A, Bocquet L, Lequeux F. 2002. Simple model for heterogeneous flows of yield stress fluids. *Physical Review E* 66:051501-12

Safouane M, Durand M, Saint Jalmes A, Langevin D, Bergeron V. 2001. Aqueous foam drainage. Role of the rheology of the foaming fluid. *Journal De Physique. IV : JP*

Saint-Jalmes A, Durian DJ. 1999. Vanishing elasticity for wet foams: Equivalence with emulsions and role of polydispersity. *Journal of Rheology* 43:1411-22

Sollich P, Lequeux F, Hébraud P, Cates ME. 1997. Rheology of Soft Glassy Materials. *Physical Review Letters* 78:2020-3

Vincent-Bonnieu S, Höhler R, Cohen-Addad S. 2006. Slow viscoelastic relaxation and aging in aqueous foam. *Europhys. Lett.* 74:533 - 9

Weaire D, Hutzler S. 1999. *The Physics of Foams*. Oxford: Oxford University Press